\documentstyle[11pt,jabIAU207_pasp,twoside,psfig,epsf]{article}
\def\edcomment#1{\iffalse\marginpar{\raggedright\sl#1\/}\else\relax\fi}
\reversemarginpar

\begin{document}
\title{The globular cluster system of M31}
\author{P. Jablonka}
\affil{D.A.E.C., Observatoire de Paris, F-92195 Meudon, France}

\begin{abstract}
This review presents  the current status of our  knowledge of M31 star
clusters.  Given  the broadness  of the subject,  I chose to  focus on
some of its aspects which are not covered by the other participants in
this conference. 

\end{abstract}

\section{Scientific issues}

Globular clusters  are studied for their intrinsic  properties as well
as for the information they give on their parent galaxy.  No less than
300 articles  address the globular  cluster system of M31.   I briefly
mention here the topics tackled along  the years and will come back in
more details to  some of them below : their  total number and specific
frequency, their stellar population: age and metallicity, the distance
to M31,  the relation between the  level of the  horizontal branch and
metallicity,  the  search   for  novae,  the  bulge/disc/halo  cluster
properties,  the  radial   distribution  of  their  properties,  their
luminosity  function, their  kinematics, dynamics,  their  true shape,
ellipticity, tidal  radii, effectif radii, the absorption  law in M31,
the mass and mass distribution  of M31, the scenario for the formation
of M31. For all these  topics, comparisons with our Galaxy and galaxies
in the Local group are usually conducted.

\section{Detection}

The clear identification and precise location of the clusters is the
first step before any further investigation. There is a very long list
of contributors, and it is nearly impossible to quote them all here.
Though I would like to mention a few historical cornerstones. Hubble
(1932) was the first to report the detection on photographic plates of
140 {\it nebulous} objects, provisionally identified as globular
clusters.  Thirty years later, in 1962, Vetesnik reported the
existence of 257 candidates.  As a matter of fact, the total number of
''confirmed'' clusters has not changed much over the years.  What has
changed is the degree of confidence that could be put in their
''confirmation''.  In 1976, Alloin, Pelat \& Bijaoui started to focus
on the inner regions of M31, and identified 8 central globular
clusters.  Ten years later, Wirth, Smarr \& Bruno (1985) increased the
list of bulge clusters up to 30 candidates.  In 1979, Hodge initiated
the search for open clusters, and proposed a list of 403 such
objects. One had to wait till 1991 for the first CCD photometry of 82
halo clusters provided by the work of Racine at the CFHT.

As to the contemporary studies of globular clusters, they are very
much based on the major study of Battistini et al. (1993). The authors
listed a grand total of 1028 globular cluster candidates, with $14~$
mag $< V < 20~$ mag , in a $3^o \times 3^o$ region (30 kpc x 30 kpc),
centered on M31.  They established their {\it best sample}, which term
is now widely used, by selecting clusters with V $\le 18~$ mag and
with B$-$V $\ge 0.58$~mag (this blue limit corresponds to the bluest
globular cluster at B$-$V = 0.52 in our Galaxy).  This sample regroups
298 objects, comprising 199 confirmed globular clusters, 72 by high
resolution imaging, 174 by spectroscopy, 47 by both methods.  Limiting
the comparison between Galactic and M31 in regions where completeness
and uncontamination apply, Battistini et al. (1993) found N$_{M31}$ /
N$_{MW}$ $\sim$ 2.5.  Recently, Barmby et al. (2000) compiled new
photometry, spectroscopy and existing data from the literature. They
inventory 435 clusters and cluster candidates, including the {\it best
sample}, 268 with 4 or more optical filter photometry, 224 with IR
photometry, 200 with velocities, 188 with spectroscopic metallicities.
M.G. Lee presents in this conference another new survey.

As to the open clusters, Williams and Hodge (2001) provide an updated
catalogue of candidates.

\section{A view from the IR}

The InfraRed offers a view of the M31 globular cluster system, the
closest to the visible. 

So far,  analyses have dealt mainly with  integrated photometry. Cohen
\& Matthews (1994)  gave a good summary of  the situation by combining
previous works  with 23  new  globular  clusters,  16 of  which   have
projected  distance less than 1.1  kpc  from M31  center.  Among these
precedent works,  Frogel et al.   (1980), Sitko (1984), Bonoli  et al.
(1992). All works combined led  to a total  sample of 84 objects.  The
mean metallicity in M31  and in our Galaxy  is the same to within 0.11
dex,   which    is  within  the   uncertainties   of  the  two  [Fe/H]
determinations.   There is  no    evidence among the   outer  globular
clusters for  a spatial  gradient of  metallicity in  the M31 globular
cluster sample, as what is observed in our Galaxy.

Stephens et  al.    present  in  these proceedings  their   HST-NICMOS
observations of M31 some metal  rich globular clusters, a study  which
opens the era of resolution of individual stars in the IR.

As  to the  spectroscopy,  Davidge (1990)  presented  spectra in  the
interval 1.5-1.8  microns and  2.0-2.4 microns of  a few  luminous M31
clusters. However  these spectra being  noisy, Davidge had to  build a
mean spectrum, from which it appeared that the CO bands are similar to
the  Galactic clusters  while CN  absorptions look  stronger.   The CN
enhancement in  M31 was  first mentioned in  the visible  (Burstein et
al. 1984).

\section{A view from the UV}

Studying  the  hot  phases of  the  stellar evolution   is  one strong
motivation  for observing globular  clusters in  the  UV, another  one
being the  understanding  of  the  connection with  the  ``UV-upturn''
population in bulges and elliptical galaxies.

Bohlin et al. (1993) observed with IUT 43 clusters; 42 were identified
in the near UV (2490 \AA) and  only 10 in the far UV (1520 \AA). Their
data suggest that the M31  clusters are close counterparts of Galactic
globulars.

Cacciari et al.  (1982) and Cowley \& Burstein (1988) obtained the
first spectra of a dozen of the brightest clusters in M31, at rather
low resolution.  Some clusters semt to exhibit residual flux below
3000 \AA, greater than that expected from the bright evolved stars in
the clusters.  Though, these pioneer observations were still at low
signal to noise and triggered discussions (Crotts et al. 1990).

The Hubble Space Telescope has opened a new era: the spectral
resolution of the UV data can now match that obtained in the visible.
As a start, Ponder et al. (1998) got HST/FOS integrated UV spectra for
4 M31 clusters.  Interestingly, the near-UV indices involving
nitrogen show an enhancement relatively to Galactic stars (disk and
halo). As in the visible, there is yet no clear explanation for this
phenomenon.

\section{A view from the X-rays}

It seems that there is a promising fied of research in the X-rays.
This domain allows a direct insight into the composition of the
globular cluster stellar population.

The  Einstein  Satellite has   brought  a   first set  of  significant
information (Battistini et  al.  1982, Trinchieri \& Fabbiano  1991).
However, ROSAT  gave a more general  view (Supper et  al.  1997, after
Primini et al.   1993), since the ROSAT  PSPC deep survey  could cover
the whole   galaxy,  with a total  integration   time about ten  times
greater than that of the Einstein observations.  In total, 396 sources
were  found in the ROSAT  PSPC first  survey  of M31,  and 27 of these
ROSAT sources were identified with globular clusters.  The fraction of
X-ray bright clusters among     the total cluster    population looked
similar in M31  and in our Galaxy.  For the two galaxies,  the maximum
luminosity  of X-ray bright  globular clusters looked also comparable.
With the second  ROSAT PSPC deep survey,  Supper et  al. (2001) update
and   confirm  these results.   However,   {\it Chandra} results, with
observation of 30 clusters, as presented by  Di Stefano et al. (2001),
seems to contradict this view.  Indeed, the authors find a higher peak
in luminosity and a larger fraction of high luminosity clusters.

At this stage, it is difficult to distinguish for which reason such
opposite conclusions can be reached. Samples, statistical analyses are
different. Further investigation seem highly needed.

\def\bsax{BeppoSAX\,}

Very    interestingly, the X-ray    technology  now allows rather good
spectroscopy and  this first  step is reported  by  Trinchieri\ et  al
(1999) with the observation of 8 clusters.   The detected sources with
\bsax\ have high X--ray luminosity (L$_X
\ge 5  \times 10^{37}$ erg  s$^{-1}$ in the  2-10 keV band).  
This suggests that they are most likely Low-Mass-X-ray Binaries.  Most
of them have similar spectrum, that can be described with a single
temperature component with kT$\sim 6-9$ keV.  Contrary to a recent
proposal, no trend shows up between of the soft X--ray properties and
the cluster metallicities.

\section{A view with the Space Telescope}

\begin{figure}
\centerline{\vbox{
\psfig{figure=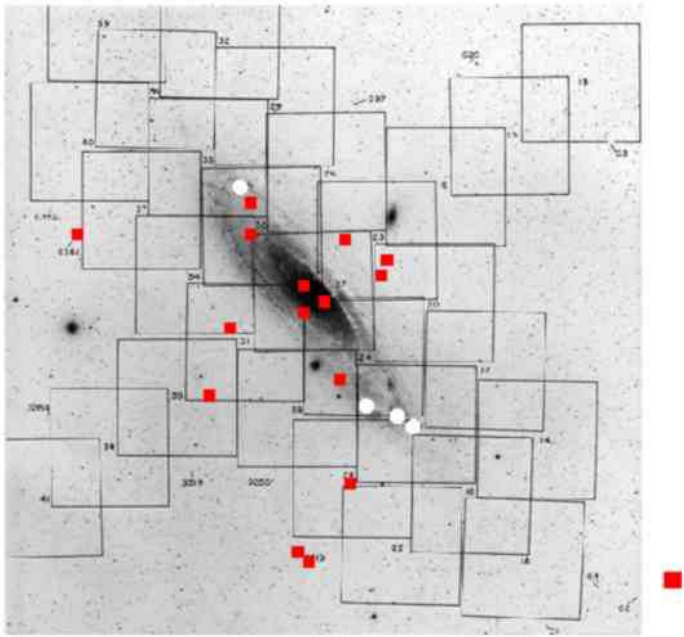,width=10.5cm,angle=0}
\caption{The M31 clusters for which color-magnitude diagrams have 
been built, at the time of Symposium. Overplotted on the M31 Atlas of
Hodge (1961), the  squares locate the observed globular clusters,
while the circles indicate the position of the observed young
clusters} }}
\end{figure}

At the time  of this  Symposium, about  twenty M31 clusters  have been
imaged   by  the Space  Telescope  with  enough spatial resolution and
photometric quality so that   their color-magnitude diagrams could  be
built. Figure 1 indicates their location in  projection over M31. Even
if  the  statistics   remain modest, halo,   disk   and  bulge cluster
populations have been sampled.

\begin{figure}
\centerline{\vbox{
\psfig{figure=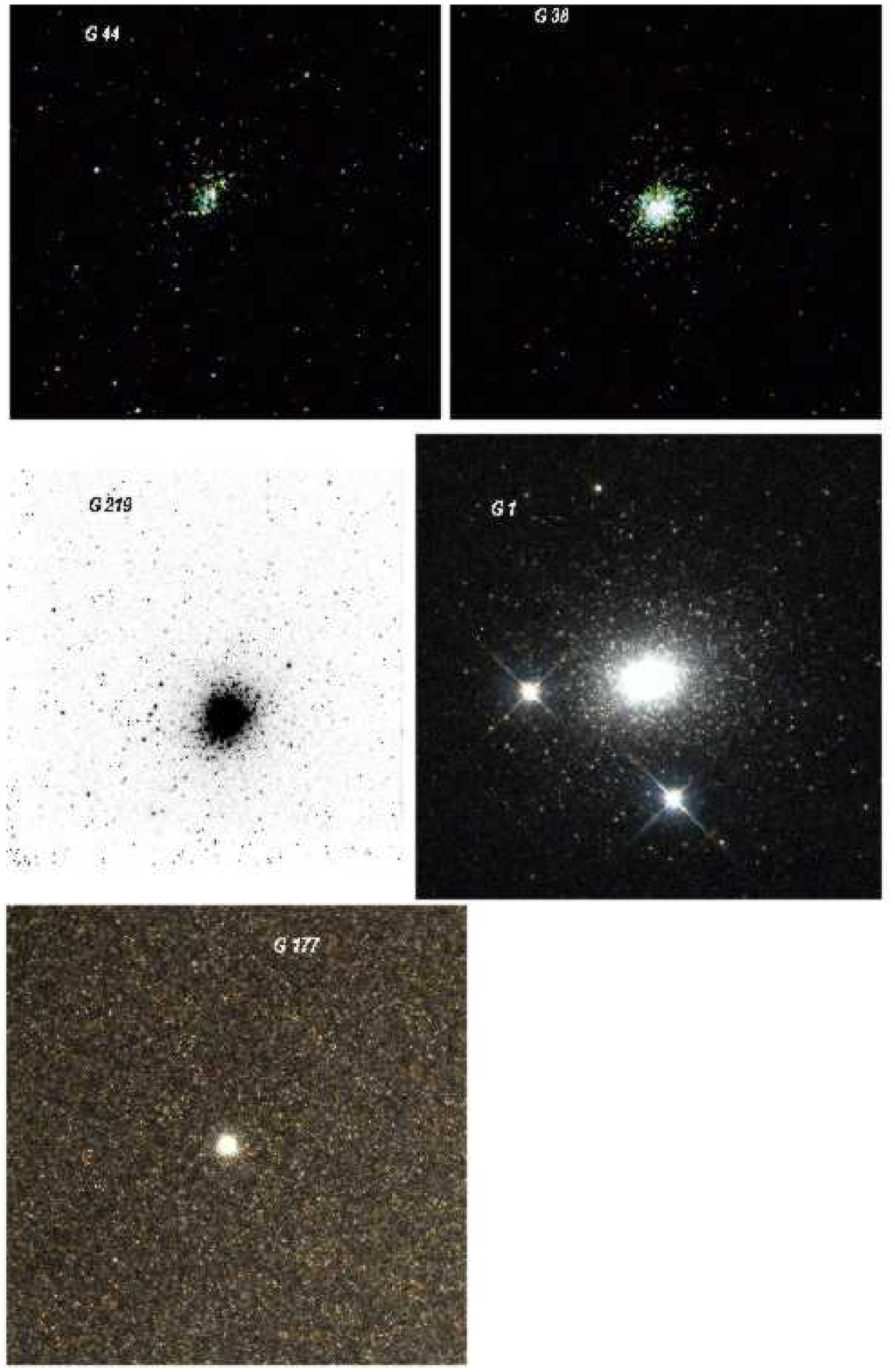,width=14cm,angle=0}
\caption{ A sample of 5 M31 globular clusters observed with HST/WFPC2:
the young clusters G44 and G38, the disk cluster G219, G1 located in
the outer halo, and the bulge cluster G177.}
}}
\end{figure}

Figure 2 displays  a variety of  cluster  morphologies. Thanks to  the
courtesy of P. Hodge and  B. Williams, the first  two upper images are
young open and compact clusters, viz. G44 and G38.  Below, on the left
hand side, thanks  to the courtesy  of C.  Cacciari, L.  Federici  and
F. Fusi Pecci, the disk globular cluster G219.  On the right side, G1,
the  most distant halo cluster.   The image is  a composite of V and I
images   (Meylan et  al.  2001).    This cluster  is  very  bright and
massive, and displays a significant flattening. Meylan details some of
its outstanding properties in these proceedings.   G177, at the bottom
of the figure, is  a bulge cluster.  It  is conspicuous that the field
surrounding this bulge  cluster is  significantly more populated  than
the  fields surrounding   the    other clusters.  Due    to  dynamical
interaction   with  the bulge, G177  has   probably lost a significant
fraction of the stars in its envelope.  Its stellar population is very
red, as would be an old and metal rich stellar system.

\subsection{An example of color-magnitude diagrams: three bulge star 
clusters}

At the distance of M31, the  best color-magnitude diagrams reach stars
about one   to two magnitudes  below  the horizontal   branch level (I
$\sim$ 24.5 mag).  See, for example, Ajhar et  al.  (1996), Fusi Pecci
et al.  (1996), Rich et al.  (1996), and Holland et  al.  (1997).  See
also B. Williams in these proceedings for the young star clusters.

As seen in  Figure 1, most clusters observed  with HST  are located in
the disk and  halo of M31.  Jablonka et  al.  (2000) have  focussed on
three   clusters  projected at   distances between  slightly  less and
slightly more  than 1~kpc from the  nucleus  of M31. The  cluster G177
seen in Figure 2 is  one of them.  Given the  extreme density of field
stars in the bulge of M31 (from 30 to  55 stars per arcsec$^2$ for V
$\leq$ 26.5  mag and I $\leq$ 24.5  mag) and the compactness  of the
clusters, we adopted  the MCS deconvolution technique (Magain, Courbin
\& Sohy,  1998).  The principle of this  method is to consider that, in
order   to  satisfy the sampling  theorem,   the deconvolution must be
conducted with a  PSF which is narrower  than the observed one. One of
its  appreciable advantages  is that  it enables  the user to  perform
photometry on the deconvolved images.  With this strategy we succeeded
in disentangling field from cluster stars, although marginally for the
innermost  cluster G198.  Comparison between   field and cluster stars
color-magnitude diagrams confirm that these  clusters were formed from
the same material as the field stars, early in the formation of M31.

\subsection{Surface-brightness profiles}

The precision  achieved  with the   HST is   such that the   questions
addressed from the  surface-brightness profiles resemble more and more
those tackled for the clusters in our Galaxy. The structural parameters
are derived (Fusi  Pecci et al.  1994  ; Grillmair et al. 1996); tidal
radii  are  computed (Cohen \& Freeman,  1991);  even  a collapse core
cluster has been identified (Bendinelli et al. 1993).

\subsection{Mass-to-Light ratios}

Obtaining  the  velocity dispersion   of   the globular clusters   and
combining  them with structural  and photometric  parameters allows to
measure the  $ M/L$  ratios,  to explore the correlations   of cluster
properties and  to compare them  with the equivalent  correlations for
the  Galactic  globular clusters.  Two  contemporary  works started to
address this aspect: Dubath \&  Grillmair (1997) and Djorgovski et al.
(1997).  The  latter, with 27 clusters,   certainly gather the largest
sample.  The authors  investigated the correlations in two  parameters
planes constructed with  the central velocity  dispersions and central
surface brightnesses   or  absolute   magnitudes, both   in V   and  K
bands. They stress  that the M31  globular clusters  smoothly continue
and extend  to higher luminosities   the trends drawn by the  Galactic
clusters.  Moreover,  at same metallicity,   Galactic and M31 clusters
have the same $M/L$ ratios.

The so far most precise dynamical study of a  globular cluster in M31,
viz.  G1, based on its surface-brightness profile and central velocity
dispersion, has  been recently published by Meylan  et  al.  (2001 and
these proceedings).

\section{Conclusion}

It clearly appears that  one will gain in  increasing the size  of the
cluster samples for which the  quality of the  data will catch up with
what is available for the clusters in our Galaxy,  as one started with
the HST.    The answer  to the  question  of  the universality  of the
cluster properties  is at this price.  Meanwhile,  the opening  of new
wavelength domains   with  good    sensitivity and    improvement   of
spatial/spectral resolutions certainly  give to the study of  globular
clusters in M31  some fresh air !  This   is becoming a  new domain of
investigation in which excitement for new insight is genuine.

\end{document}